# Fingerpad Contact Evolution Under Electrovibration

**Omer Sirin[1], Allan Barrea[2,3], Philippe Lefèvre[2,3], Jean-Louis Thonnard[2,3], Cagatay Basdogan[1]**

[1]Koc University, College of Engineering, Robotics and Mechatronics Laboratory, 34450, Istanbul, Turkey
{osirin13, cbasdogan}@ku.edu.tr
[2]Institute of Neuroscience (IoNS), [3]Institute of Information and Communication Technologies, Electronics and Applied Mathematics (ICTEAM), Université catholique de Louvain, Brussels and Louvain-la-Neuve, Belgium {allan.barrea, philippe.lefevre, jean-louis.thonnard}@uclouvain.be

**Abstract**
Displaying tactile feedback through a touchscreen via electrovibration has many potential applications in mobile devices, consumer electronics, home appliances, and automotive industry though our knowledge and understanding on the underlying contact mechanics is very limited. An experimental study was conducted to investigate the contact evolution between the human finger and a touchscreen under electrovibration using a robotic set-up and an imaging system. The results show that the effect of electrovibration is only present during full slip but not before slip. Hence, coefficient of friction increases under electrovibration as expected during full slip, but the apparent contact area is significantly smaller during full slip when compared to that of no electrovibration condition. It is suggested that the main cause of the increase in friction during full slip is due to an increase in real contact area and the reduction in apparent area is due to stiffening of the finger skin in tangential direction.

**Introduction**
Currently, there are two major techniques, namely ultrasonic and electrostatic actuation, for displaying tactile feedback through a haptic surface. In both techniques, the aim is to modulate the frictional forces between finger pad and the surface, though their working principles are different. When a haptic surface is actuated at an ultrasonic resonance frequency, friction is reduced between the surface and finger [1- 5]. On the contrary, in electrostatic actuation, a voltage is applied to the conductive layer of a surface capacitive touchscreen to generate an electroadhesive force between its surface and a finger sliding on it, which leads to an increase in frictional forces applied to the finger opposite to the direction of movement [6-9]. A surface capacitive touch screen has a single layer of conductive ITO (indium tin oxide) placed above a glass substrate and covered by an insulator layer. When a voltage is applied to this conductive layer, a uniform and attractive electrostatic force field is generated between finger pad and the touch screen in the direction normal to its surface. Although the magnitude of this electrostatic force is small relative to the normal force applied by the finger, it results in a perceivable frictional force in tangential direction when the finger pad slides on the touch screen. This frictional force can be modulated to generate different tactile effects by altering the amplitude, frequency, and phase of the voltage signal applied to the conductive layer of the touch screen. While the technology for generating tactile feedback on a touch screen via electrovibration is already in place and straightforward to implement, our knowledge on the underlying contact mechanics and tactile perception are highly limited [10]. This is not surprising since the contact interactions between human skin and a counter surface is already highly complex to investigate even without electrovibration [11-13]. In the case of electrovibration in particular, the exact mechanism leading to an increase in tangential frictional forces is still not known completely. To estimate the electrostatic forces between finger pad and touch screen, models based on parallel-plate capacitor theory have been proposed (see the summary in Vodlak et al. [14]). These models assume a constant air gap between two surfaces and ignore the rounded shape of the finger pad and the asperities on its surface. On the other hand, to estimate the frictional forces in tangential direction during sliding, the Coulomb model of friction utilizing a constant coefficient of dynamic friction between finger pad and touch screen has been used so far. The increase in tangential force has been explained by simply adding the force due to electrostatic attraction to the normal force applied by the finger, $F_t=\mu(F_n+F_e)$, where $\mu$ is the friction coefficient, $F_t$, $F_n$, and $F_e$ represent the tangential, normal, and electrostatic forces, respectively. However, friction of human skin against smooth surfaces is governed by adhesion model of friction [15-17], which depends on interfacial shear strength and real area of contact. The frictional force due to adhesion is given by $F_t = \tau A_r$, where $\tau$ is the interfacial shear strength and $A_r$ is the real area of contact. It is highly difficult to measure or estimate the real area of contact, which varies nonlinearly with the normal force. In earlier studies, the real contact area has been taken as the contact area of finger ridges, $A_{ridge}$, which is smaller than the apparent contact area, $A_{apparent}$, estimated by the fingerprint images, using the boundaries of finger pad in contact with surface [18]. On the other hand, if the Persson's contact theory is considered (also see the review of other contact theories in the same study [19]), the real area of contact ($A_r$) is, in fact, much smaller than the contact area of finger ridges ($A_{ridge}$) since only the finger asperities at finer resolution make contacts with the surface. For the interfacial shear strength of skin, Adams et al. [20] adopted a model from an early study by Bowden and Tabor [21], which suggests that $\tau$ increases linearly with the mean contact pressure as $\tau = \tau_0 + \alpha p_r$, where $\tau_0$ is the intrinsic shear strength, $\alpha$ is the pressure coefficient, and $p_r = F_n/A_r$ is the real contact pressure. Hence, the adhesive friction can be expressed as $F_t = \tau_0 A_r + \alpha F_n$. Briscoe and Tabor [22] suggested that for polymers, the friction coefficient tends to the pressure coefficient $\alpha$ when the contact pressure ($p_r$) is higher than $10^9$ Pa. However, typical contact pressure between finger pad and touchscreen is much lower than this value [23]. Hence, under the assumption that $\alpha$ does not vary significantly with and without electrovibration (EV) for the same normal load and sliding velocity, then the main factors affecting the adhesive friction are intrinsic shear strength ($\tau_0$) and real contact area ($A_r$).

In this paper, the increase in adhesive frictional force due to electrovibration is investigated before and after slip. We hypothesize that the effect of electrovibration is only present during full slip but not before slip. This hypothesis was supported by presenting the results of experiments conducted with six human subjects at a constant finger sliding velocity under three different normal loads. We also present the adverse effect of fingertip skin moisture on electrovibration.

## Methods
### Set-up
The experiment was performed with the help of a robotic system to maintain a constant normal contact force between the finger and a touch screen while ensuring a constant sliding speed. A surface capacitive touch screen (SCT3250, 3M Inc.) was cut in dimensions of 10x10 cm and attached to the end-effector of the robot (HS4535G, Denso Inc.) along with two force sensors (Mini-40, ATI Inc.) to measure contact forces between the index finger of subject and touch screen in tangential and normal directions. The measured normal force was fed back to a PID controller to keep the normal force applied by the robot constant. The tangential force was measured with and without EV. To activate electrovibration, the touch screen was stimulated with a voltage signal generated by a signal generator (Model 182A, Wavetek) and augmented by an amplifier (E-413, Physik Instrumente). A conductive copper tape (3M Inc.) was wrapped around the edges of touchscreen to transmit voltage to its conductive layer. The subject's index finger was fixed in a support to maintain a constant angle of contact (20 degrees) with the touch screen. The end-effector position was servo-controlled to translate the touch screen at a constant speed during the experiments. The fingerprint images were acquired at 200 Hz by an imaging system that includes a light source, mirrors, and a high-resolution camera (EoSens, MC1362, 1280x1024 pixels, Mikrotron Inc.). The details of the set-up and the data collection methods are available in Delhaye et al. [24].

**Participants**
The experiment was conducted with 6 adult subjects having an average age of 32.3 years (SD: 15.1). The subjects used the index finger of their right hand during the experiments. All subjects provided written informed consent to undergo the procedure, which was approved by the ethics committee at Université catholique de Louvain. The investigation conformed to the principles of the Declaration of Helsinki and the experiment was performed in accordance with relevant guidelines and regulations.

**Experimental Procedure**
Before the experiment, the subjects washed their hands with soap, rinsed with water, and dried them in room temperature before the experimentation. Moreover, the touch screen was cleaned with ethanol. The subjects were asked to wear a ground strap on their stationary wrist.
During the experiments, the subjects placed their index finger into the support to make a constant angle of contact with the touch screen. They kept their finger on the screen for 5 seconds with the help of the robot so that the normal force could be adjusted and some moisture accumulated at the interface. Then, the robot was commanded to move the touch screen with a constant velocity of 20 mm/s while keeping the normal force constant. The movement of the touch screen was along a single direction (radial) and from right to left. To ensure a good contrast in captured images of the contact area, a small amount of liquid Vaseline oil was applied to the imaged fingertip before starting each experiment. Then, the tangential and normal forces were recorded as a function of displacement with and without EV. To activate electrovibration, a sinusoidal voltage, at a frequency of 125 Hz with an amplitude of 200 Vpp, was applied to the touch screen. The experiments were conducted for 3 different normal forces (0.5, 1, and 2 N). Each subject performed 30 trials (3 normal forces x 1 finger velocity x 2 experimental conditions x 5 repetitions). The moisture level of each subject's fingertip skin was measured by a Corneometer (CM 825, Courage - Khazaka Electronic) before the first and after the last trial of each experiment three times for each normal force. Hence, a total of 18 data points was collected on moisture level for each subject (measurements performed before the first and after the last trial x 3 repetitions x 3 normal forces).

# Results

Before Full Slip: To quantify the proportion of contact area slipping against the glass, feature points were sampled within the apparent contact area and tracked using optical flow. The onset of full slip was defined as the moment when all tracked points were moving relative to the glass (the details of this approach are available in Delhaye et al. [24]). This corresponds to a stick ratio of zero (SR=0), where SR is defined as the proportion of the contact moving with the glass, i.e. the ratio of the no-slip area to the initial apparent contact area. In this formalism, SR=1 for a stuck contact and decreases down to SR=0 during the transition from stick to slip. Figure 1a shows the evolution of SR for one subject as a function of the tangential force with/without EV for a normal force of 2 N. In order analyze the data, we used tests that are "robust" to the assumption of normality, such as Analysis of Variance (ANOVA) and t-test. A three-way analysis of variance (ANOVA) was performed with repeated measures on the tangential force with and without EV, normal force ($F_n$ = 0.5 N, 1 N, 2 N), and the stick ratio (SR = 1, 0.75, 0.50, 0.25) as the main factors. The difference between the tangential forces with and without EV was not significant (F (1,5) = 1.54, p = 0.27) though the normal force and the stick ratio significantly affected the tangential force, as expected.

Figure 1b depicts the initial (no motion) apparent contact area ($A_0$) of subjects' finger pad for different normal forces (0.5 N, 1 N, 2 N) with and without EV. The results were analyzed by two-way ANOVA with repeated measures. No significant difference was observed between the initial apparent contact areas with and without EV (F (1,5) = 0.254, p = 0.635). However, the effect of normal force on the initial apparent contact area was significant (F (2,10) = 146.529, p < 0.001), as expected. There was no statistically significant interaction between electrovibration and normal force.

During Full Slip: Figure 2a shows the normalized apparent contact area (the ratio of apparent contact area to the initial apparent contact area; $A_{apparent}/A_0$) as a function of finger displacement for each subject. The normalization helps to compare the change in apparent contact area across subjects. The apparent contact area of each subject reached to a steady-state value during full slip with and without EV. Figure 2c displays the fingerprint images of subject 4 (S4). The steady-state value of the apparent contact area ($A_{steady-state}$) was calculated by using the image data corresponding to the displacement of the last 5 mm and then the percent reduction in the steady-state value of the apparent contact area ($A_{steady-state}$) with respect to the initial apparent contact area ($A_0$) was reported in Figure 2b. The percent reduction in $A_{steady-state}$ (with respect to $A_0$) was larger with EV. The results were analyzed using two-way ANOVA with repeated measures. The ANOVA results showed that both electrovibration (F (1,5) = 60.71, p = 0.001) and normal force (F (2,10) = 16.793, p = 0.001) had a significant effect on the percent change in the steady-state value of the apparent contact area ($A_{steady-state}$) with respect to the initial apparent contact area ($A_0$). There was no statistically significant interaction between electrovibration and normal force.

Figure 3a shows the coefficient of friction (μ) as a function of finger displacement for each subject. The coefficient of friction reached to a steady-state value ($μ_{steady-state}$) with and without EV. Figure 3b shows the mean values of steady-state friction coefficient, $μ_{steady-state}$, (estimated from the friction data corresponding to the finger displacement of the last 5 mm) for 3 different normal forces. The steady-state friction coefficient, $μ_{steady-state}$, increased under electrovibration for all normal forces. The results were analyzed using two-way ANOVA with repeated measures. The ANOVA results showed that both electrovibration (F (1,5) = 21.659, p = 0.007) and normal force (F (2,10) = 34.413, p<0.001) had a significant effect on the steady-state friction coefficient, $μ_{steady-state}$. There was a statistically significant interaction between electrovibration and normal force (F (2,10) = 11.248, p = 0.003). The effect of electrovibration on the steady-state friction coefficient, $\mu_{steady-state}$, for each normal force was further analyzed by paired t-tests and found significant ($p < 0.05$).

Effect of Moisture: The coefficient of friction on smooth glass has been shown to vary due to fingertip skin moisture [25]. The touch screen used in our study has a smooth surface with a surface roughness that was measured to vary between 40-220 nm. The subjects were ordered according to their finger pad moisture levels as S1 = 34.1±2.1, S2 = 39.1±8.8, S3 = 59.0±11.9, S4 = 72.7±10.0, S5 = 80.3±8.1, S6 = 115.5±1.4. Since the fingertip skin moisture level of each subject was different, we investigated the relative effect of moisture on the coefficient of friction with and without EV, using Tactile Friction Contrast (TFC), a metric proposed by Cornuault et al. [26]. As shown in Figure 4, TFC is smaller for higher levels of fingertip skin moisture, suggesting an adverse effect of moisture on the capacity of electrovibration to modulate friction for the three levels of normal force (panels a, b and c of Figure 4). However, a moisture-controlled study with higher number of subjects is necessary to make an affirmative conclusion.

## Discussion

The results showed that the effect of electrovibration was only present during full slip but not before slip. We arrived to this conclusion based on the tangential forces and apparent contact areas measured with and without EV. During the transition from stick to full slip, the transient behavior of stick ratio (SR) as a function of tangential force was similar with and without EV under different normal forces (Figure 1a). In addition, our results also showed that the apparent contact areas before the onset of full slip were similar with and without EV (Figure 2a) while they both reduced with respect to the initial apparent contact area (Figure 2b). On other hand, Sahli et al. observed that the finger ridge area decreases under increasing shear, well before the onset of sliding during contact interactions between human finger and a glass surface [13]. Delhaye at al. also investigated the evolution of the apparent area of contact between finger and glass during stick-to-slip transition on a smooth glass surface without electrovibration [24]. They estimated the intrinsic shear strength $\tau_0$ using the minimal tangential force causing a partial slip (see the flat section of curves in Fig. 1a for SR = 1). They observed that this force increases with normal force but decreases with speed. Since the minimal tangential force (Figure 1c) and the apparent contact area before slip did not change significantly with EV in our study, we argue that the values of intrinsic shear strength were similar with and without EV (note that the intrinsic interfacial shear strength estimated from the no-slip region is also applicable to the slip region, as argued by Adams et al. [20]). We should also point out that the finger moisture plays an important role in contact interactions as reported in this study and the moisture accumulating in the interfacial gap before the slip might have reduced the effect of electrovibration. In fact, a recent study by Shultz et al. [27] showed that the electrical impedance of the interfacial gap is significantly lower for the stationary finger compared to that of the sliding finger under electrovibration, suggesting that the role of moisture is reduced during sliding.

Once the stick-to-slip transition was over and the slip started, the apparent contact area was reduced significantly under electrovibration. There was a clear difference in the apparent contact areas between the conditions with and without EV for all tested subjects (Figure 2a). The increase in frictional forces with EV led to further reduction in apparent contact area (Figure 2b). Earlier studies performed on smooth glass (without EV) also reported a reduction in apparent contact area when the tangential force was increased [24, 28]. The authors explained this phenomenon by nonlinear stiffening of the skin in tangential direction. In our study, the tangential frictional force (and hence the coefficient of friction) also increased under electrovibration, as observed in the earlier studies. We argue that the increase in friction with EV was due to an increase in real contact area. First of all, a decrease in coefficient of friction was observed when the normal force was increased, which suggested that load-independent Coulomb friction model was not applicable, as reported in earlier studies [16]. Instead, if the adhesive friction model of Adams et al. [20] is adopted, the coefficient of friction becomes a function of normal load, intrinsic shear strength, pressure coefficient, and the real contact area. Assuming that the intrinsic shear strength does not change with EV and the pressure coefficient is effective for large contact pressures only, then the main cause for the change in frictional force is likely due to an increase in real contact area since the normal force and finger velocity were controlled and maintained constant during our experiments. We argue that the increase in real contact area is due to the electroadhesive forces, causing the finger asperities make more contacts with the surface of the touchscreen at microscopic scale. A support for our claim on the effect of real contact area comes from Persson's sliding friction theory [19], which suggests that the real contact area for elastic surfaces (such as human finger [29, 30]) having different length scales is significantly larger than it was originally thought for increasing normal load. A further and stronger support was recently provided by Sirin et al. [23] and Ayyildiz et al. [31]. They investigated the sliding friction by electrovibration using (i) a mean field theory based on multiscale contact mechanics, (ii) a full-scale computational contact mechanics study, and (iii) experiments performed on a custom-made tribometer. The results of this integrated approach showed that the electroadhesion causes an increase in the real contact area at the microscopic level, leading to an increase in the tangential frictional force. Hence, under electrovibration, microscopic asperities of fingertip skin make more contacts with the touch screen during sliding, leading to an increase in real contact area. As shown in previous work [4], each asperity making microscopic contact can further support adhesive shear load proportional to its own contact area, increasing the overall magnitude of tangential forces shearing those contacts.

We also observed that electrovibration was adversely affected by fingertip skin moisture, which has not been reported before. The electrovibration effect was less prominent for subjects having higher levels of fingertip skin moisture (Figure 4). For example, the skin moisture level of subject 6 (S6) was higher than those of the other subjects. This resulted in his coefficient of friction be higher than those of the other subjects, but the relative change in his coefficient of friction due to electrovibration was less than those of the other subjects. Earlier studies performed on smooth glass surfaces (without EV) showed that the coefficient of friction increases when there is a thin layer of water between finger and glass surface [20]. This was explained by the softening of the finger, also known as plasticization, which results in an increase in contact area and thus the tangential frictional force. In addition to the finger softening due to moisture, the air gap between finger and the surface is filled by water particles, causing a reduction in the electrostatic force in our case.

In our experiments, oil was applied to the subjects to ensure a good contrast in captured video images. As reported in the earlier studies [32, 33], oil reduces the friction coefficient at the interface. However, it was not possible to record the finger images without any use of oil in our system since the touchscreen used in our study is less transparent than the standard glass surface used in the earlier studies [24]. However, in order to minimize the potential effects of oil in our study, we made relative comparisons between the conditions with and without EV in our analyses by using metrics emphasizing the contrast rather than the absolute values.

## Conclusion

Displaying tactile effects through a touch screen has several potential applications in consumer electronics, home appliances, and mobile devices. One of the approaches in this regard is electrovibration. The contact mechanics behind the electrovibration is currently under investigation. In this study, the cause of an increase in tangential frictional forces under electrovibration was investigated. For this purpose, an experimental study was conducted. The frictional forces acting on subjects' index finger pad and its apparent contact area were measured at a constant sliding speed during transition from stick to full slip for different normal loads with and without EV. The results suggest that although the contact evolution with and without EV are quite similar before full slip, they are distinct during sliding. In particular, an increase in the coefficient of friction was observed during sliding as expected, but surprisingly a larger decrease in apparent contact area was observed under electrovibration. Based on the adhesive friction model suggested in the literature [21, 20], we proposed that the increase in friction force under electrovibration is due to an increase in the real contact area.

**Data accessibility**. This article has no additional data.
**Authors' contributions.** O.S. A.B., P.L., J-L. T., and C.B. conceived the experiments; O.S. and A.B. conducted the experiments; O.S. A.B., P.L., J-L. T., and C.B. analyzed the results; C.B. and O.S. wrote the manuscript; O.S. A.B., P.L., J-L. T., and C.B. reviewed and revised the manuscript.
**Competing interests.** The authors declare no competing interests.
**Funding.** O.S. acknowledges the scholarship provided by TUBITAK through student fellowship program BIDEB–2211. CB acknowledges the financial support provided by TUBITAK under the contract number of 117E954.

# Figure Captions

**Fig. 1.** **(a)** The evolution of the stick ratio of one subject (S2) as a function of tangential force for a normal force of 2N. The solid lines represent the average of 5 trials while the shaded areas around them represent the standard deviations **(b, c)** Mean values (average of 6 subjects) of the initial apparent contact area and minimal tangential force to start a slip as a function of normal force. The error bars represent the standard errors of the means.

**Fig. 2. (a)** Normalized apparent contact area of each subject as a function of finger displacement for a normal force of 2N. Subjects are ordered according to their moisture level from S1 having the lowest level of moisture to S6 having the highest level of moisture. The thin lines are the individual trials while the thick lines represent the average of 5 trials. The vertical dashed lines mark the instant of full slip (SR = 0). **(b)**

Mean values (average of 6 subjects) of the percent reduction in the steady-state value of the apparent contact area ($A_{steady-state}$) with respect to the initial apparent contact area ($A_0$) for 3 different normal forces. The error bars represent the standard errors of the means. **(c)** Fingerprint images of subject 4 (S4).

**Fig. 3. (a)** The coefficient of friction ($\mu$) of each subject as a function of finger displacement for a normal force of 2N (note that the subjects' finger was stationary and the touch screen was moved along a single direction, radial, and from right to left). Subjects are ordered according to their moisture level from S1 having the lowest level of moisture to S6 having the highest level of moisture. The thin lines are the individual trials while the thick lines represent the average of 5 trials. **(b)** Mean values (average of 6 subjects) of the steady-state friction coefficient ($\mu_{steady-state}$) for 3 different normal forces.

**Fig. 4.** The tactile friction contrast (TFC) as a function of finger displacement of each subject having a different level of fingertip skin moisture for 3 different normal forces **(a)** 0.5 N, **(b)** 1 N, **(c)** 2 N. Subjects are ordered according to their moisture level from S1 having the lowest level of moisture to S6 having the highest level of moisture.